\documentclass{PoS}
\usepackage[latin9]{inputenc}
\usepackage{units}
\usepackage{amsmath}

\makeatletter

\title{Perturbative running of the twisted Yang-Mills coupling in the gradient flow scheme}

\ShortTitle{Perturbative running of the twisted Yang-Mills coupling in the gradient flow scheme}

\author{\speaker{Eduardo I. Bribian} and Margarita García Pérez%
\\
       Instituto de Física Teórica UAM-CSIC,\\
       Nicolás Cabrera 13-15,
       Universidad Autónoma de Madrid, E-28049-Madrid,Spain\\
       E-mail: \email{e.i.bribian@csic.es, margarita.garcia@uam.es}}

\abstract{We report on our ongoing computation of the perturbative running of the Yang-Mills coupling using gradient flow techniques.\
           In particular, we use the gradient flow method with twisted boundary conditions to perform a perturbative expansion of\
   the expectation value of the Yang-Mills energy density up to fourth order in the coupling at finite flow time. We regularise the resulting\
   integrals using dimensional regularisation, and reproduce the universal coefficient of the $1/\epsilon$ term in the relation between bare and
   renormalised couplings. 
The computation of the finite part leading to a determination of the $\Lambda$ parameter in this scheme is 
underway. }

\FullConference{34th annual International Symposium on Lattice Field Theory\\
24-30 July 2016\\
University of Southampton, UK}

\makeatother

\newcommand{\be}{\begin{equation}}
\newcommand{\ee}{\end{equation}}
\newcommand{\ba}{\begin{array}}
\newcommand{\ea}{\end{array}}
\newcommand{\baa}{\begin{array}}
\newcommand{\eaa}{\end{array}}
\newcommand{\bea}{\begin{eqnarray}}
\newcommand{\eea}{\end{eqnarray}}
\newcommand{\half}{\frac{1}{2}}

\newcommand{\lef}{\tilde l}
\newcommand{\thetat}{\tilde \theta}
\newcommand{\vef}{V_{\rm eff}}
\newcommand{\ec}{{\cal E}}
\newcommand{\hc}{{\hat c}}

\begin{document}

\section{Introduction}

In this talk we report on our ongoing calculation of the perturbative running coupling of the $SU(N)$ gauge theory 
in the twisted gradient flow scheme (henceforth TGF) \cite{Ramos:2014kla}. The purpose of the project is twofold. First, we aim 
at a determination of the ratio of $\Lambda$ parameters between this scheme and the $\overline{\rm MS}$ scheme of dimensional regularisation, a quantity
that can be exactly determined at one-loop in perturbation theory (see \cite{okawa06} for a recent non-perturbative determination). 
Second, we want to analyse, through the running coupling, the interplay between volume and group degrees of freedom in the gauge theory. 
This comes along with the idea of volume independence, see e.g. \cite{GonzalezArroyo:1982hz}-\cite{Unsal:2008ch}, 
which has been around in twisted gauge theories ever since 
the formulation of the Twisted Eguchi-Kawai model \cite{GonzalezArroyo:1982hz}; a single site formulation of $SU(N)$ 
lattice gauge theory reproducing in the large $N$ limit the infinite volume gauge theory. The idea has since been generalised 
into the hypothesis that, in a $SU(N)$ theory on a $d$-dimensional twisted torus, the physical size and the size of the gauge group 
always appear combined into an effective length $\lef$, to be defined below \cite{Perez:2014sqa,Perez:2013dra}. 
In this work,  we will use $\lef$ to set the scale of the running coupling \cite{Perez:2014isa} 
and will test whether or not volume independence holds.

\section{The Twisted Gradient Flow Scheme}

In this context, we used the TGF scheme \cite{Ramos:2014kla} to obtain 
the running of the renormalised \textquoteleft t Hooft coupling $\lambda=g^{2}N$, with $g$ denoting the Yang-Mills coupling.

We start by formulating the $SU(N)$ gauge theory on a $d$-dimensional torus with twisted boundary conditions (TBC). We shall 
restrain ourselves to the case of the symmetric twist: $n_{\mu\nu}=  \epsilon_{\mu\nu} k l_g$, 
where $k$ and $l_g$ are two coprime integers and $\epsilon_{\mu\nu} = \theta(\nu - \mu) -  \theta(\mu - \nu)$. The gauge length $l_g  =N^{\frac{2}{d_{t}}}$ is given in terms of both $N$
and the number $d_{t}$ of dimensions with TBC.
We will consider a torus of period length $l$ in the twisted directions, and $\lef=l_g  l$, in the non-twisted ones~\cite{Keegan:2015lva}, for a torus volume $V=l^{d_{t}} \lef^{d-d_{t}}$.
Then, the effective volume controlling finite volume effects is $\vef= \lef^d$.

The gradient flow \cite{Narayanan:2006rf}-\cite{Lohmayer:2011si} works by defining an additional time dimension $t$ called flow time, and
introducing a new flow field $B_{\mu}(x,t)$ following the so-called flow equations:
\be
\partial_{t}B_{\nu}\left(x,t\right)=D_{\mu}G_{\mu\nu}\left(x,t\right)
\ee
with the initial condition that $B_{\mu}(x,0)$ matches the original $A_{\mu}(x)$ Yang-Mills gauge field.
$D_\mu$ and $G_{\mu \nu} $ respectively denote the flow field's covariant derivative and field strength.

The action density at finite flow time, $E(t)$, is a renormalised quantity allowing to define a renormalised coupling, which in the TGF 
scheme runs with the size of the twisted torus \cite{Ramos:2014kla}. We will follow the proposal in Ref. \cite{Perez:2014isa} and set the running
scale in terms of $\lef$. The coupling is then:
\be
\lambda_{\rm TGF}(\lef)=\mathcal{N}^{-1}(c)\left.\frac{t^{2}\langle E(t)\rangle }{N}\right|_{t=\frac{1}{8}c^{2}\lef^{2}}\qquad\qquad E(t)=\half \text{Tr }G_{\mu\nu}(x,t)G^{\mu\nu}(x,t)
\ee
Here $\mathcal{N}$ denotes a normalisation constant obtained by matching the coupling to the tree level bare one, and given in terms of the Jacobi theta function $\theta_{3}$:
\begin{equation}
\mathcal{N}=\frac{c^{4}\left(d-1\right)}{128\tilde{l}^{d-4}}\theta_{3}^{d-d_{t}}\left(0,i\pi c^{2}\right)\left\{ \theta_{3}^{d_{t}}\left(0,i\pi c^{2}\right)-\theta_{3}^{d_{t}}\left(0,i\pi c^{2}l_g^2\right)\right\};\qquad\theta_{3}\left(0,is\right)=\underset{m\in\mathbb{Z}}{\sum}e^{-\pi sm^{2}}
\end{equation}

\section{Perturbative expansion of  $\langle E\rangle$}

We are now ready to derive the perturbative expansion of $\langle E\rangle$. We have followed the strategy introduced in
Ref. \cite{Luscher:2010iy}, adapted to the finite volume twisted setup.   
We expand the flow fields in powers of the coupling and solve the modified flow equation order by order in perturbation theory;
\be
\partial_{t}B_{\nu}(x,t)=D_{\mu}G_{\mu\nu}(x,t)+D_{\nu}\partial_{\mu}B_{\mu}(x,t);\qquad
B_{\mu}(x,t)=\sum_k g_0^k(t) B_\mu^{(k)}(x,t).
\ee
 This is most easily done in momentum space, which in our case reads: 
\be
B_\mu^{(k)}(x,t)=\frac{1}{\sqrt{V}}{\sum^{\prime}_q} e^{iqx}B_{\mu}^{(k)}(q,t)\hat{\Gamma}(q).
\ee
All momenta are quantized in units of $\tilde l$. The prime indicates the exclusion from the sum of momenta for which $m_\mu=0\,(\text{mod }l_{g})$ in all twisted directions.
The momentum dependent basis $\hat{\Gamma}(q)$  is characterised by the following commutation relation and structure constants:
\be
\left[\hat{\Gamma}(p),\hat{\Gamma}(q)\right]=i F(p,q,-p-q)\hat{\Gamma}(p+q);
\qquad F(p,q,-p-q)=-\sqrt{\frac{2}{N}}\sin\left(\frac{1}{2}\theta_{\mu\nu}p_{\mu}q_{\nu}\right).
\ee
We defined an auxiliary tensor $2\pi\theta_{\mu\nu}=\thetat\lef^2\tilde{\epsilon}_{\mu\nu} $, where 
$\tilde{\epsilon}_{\mu\nu}$ satisfies $\tilde{\epsilon}_{\mu\nu}\epsilon_{\nu\lambda}=\delta_{\mu\lambda}$,
and $\thetat= \bar k /l_g$, with $\bar{k}$ given by the twist: $k\bar{k}=1\,(\text{mod }l_{g})$.

Expanding the action density in terms of the $B_{\mu}^{(k)}$ fields up to order $g_{0}^{4}$, we obtained several terms such as, among others (following the notation in \cite{Luscher:2010iy}):
\bea
\ec_{0}&=&\frac{g_0^2}{2 N V}\sum^{\prime}_q (q^{2}\delta_{\mu\nu}-q_{\mu}q_{\nu})\left\langle B_\mu^{(1)}(-q)B_\nu^{(1)}(q)\right\rangle,  \\
\ec_{1}&=&-\frac{g_0^3}{N V^{\nicefrac{3}{2}}} \underset{p_1,p_2,p_3}{\sum^{\prime}}F(p_1,p_2,p_3)\delta(p_1+p_2+p_3) i p_{1\mu}\left\langle B_\nu^{(1)}
(p_1)B_{\mu}^{(1)}(p_2)B_{\nu}^{(1)}(p_3)\right\rangle, \\
\ec_{2}&=&\frac{g_0^3}{ N V}\sum^{\prime}_q (q^2\delta_{\mu\nu}-q_{\mu}q_{\nu})\left\langle B_{\mu}^{(1)}(-q)B_{\nu}^{(2)}(q)\right\rangle.
\eea
Introducing the solutions of the flow equations into these expressions, defining  $q=p+r$ and after some algebra we are left with:
\bea
\ec_{0}&=&\frac{\lambda_0}{2\vef}\sum^{\prime}_q e^{-2tq^2} \left ((d-1)+\frac{\lambda_0}{2\vef} \, \sum_{p}\, \frac{N F^{2}(r,p,-q)} {p^2 q^2 r^2} 
\left[(3d-2)q^2- 2(2-d)^2 p^2\right]\!\!\right),\\
\ec_{1}&=&\frac{3(1-d)\lambda_0^2 }{2 \vef^2}\, \sum_{r,p} \, \frac{N F^2(r,p,-q)}{p^2 r^2} e^{-t (p^2+q^2+r^2)} ,\\
\ec_{2}&=& \frac{\lambda_0^2}{\vef^2} \,  \int_0^t dx \, \sum_{r,p} \, \frac{N F^2(r,p,-q)}{p^2 q^2 r^2}  \, e^{-(2t-x) q^2 - x  (r^2+p^2)} \\
&&
\left \{(d-1) q^2 (r^2+p^2+q^2) + 2 (d-2) (p^2 r^2 - (p\cdot r)^2)\right \}. \nonumber 
\eea
The Schwinger representation will be used for the propagators, and we will focus on one of the integrals contributing to $\ec_{2}$ to illustrate our computation procedure:
\be
\label{eq.i2}
I(t)=  \frac{1}{\vef^2} \int_0^t dx \int_0^\infty dz \sum_{r,p} N F^{2}(r,p,-q) e^{-(2t-x) q^2 - x  (r^2+p^2)- z r^2} .
\ee
We fix $t$ to $c^2 \lef^2 /8$, as our objective is to determine the running coupling at that scale. After rescaling some variables and using the quantisation of momenta in units of $\lef$, we arrive at:
\be
I(t_0)=  \frac{\hat c^2}{ 16 \pi^2 \lef^{2d-4} } \int_0^1 dx \int_0^\infty dz \sum_{m,n \in \mathbb{Z}^{d}} e^{- \pi \hat c (2 m^2 + (z+2x) n^2 - 2x m\cdot n)  }
\Big (1-\cos(2 \pi \thetat m\cdot \tilde{\epsilon} \cdot n)\Big ) ,
\ee
where $\hat c = \pi c^2 /2$.
A more compact way of writing the momentum sums is possible using Siegel theta functions:
\begin{equation}
\Theta\left(0\vert iA(s,u,v,\thetat)\right)=\underset{M\in\mathbb{Z}^{2d}}{\sum}\exp\left\{ -\pi M^{t}A\left(s,u,v,\tilde{\theta}\right)M\right\} .
\end{equation}
Where we defined $M^t = (m,n)$ and the following matrix:
\begin{equation}
A\left(s,u,v,\tilde{\theta}\right)=\begin{pmatrix}s\hat c\, \mathbb{I}_{d} & v\hat c \, \mathbb{I}_{d}+i\tilde{\theta}\tilde{\epsilon}\\
v \hat c\,  \mathbb{I}_{d}-i\tilde{\theta}\tilde{\epsilon} & u \hat c\,  \mathbb{I}_{d}
\end{pmatrix}
\end{equation}
These theta functions always appear in the same combination, so we
reabsorbed them into:
\begin{equation}
F_{c}\left(s,u,v,\tilde{\theta}\right)=\frac{\hat c^2 }{16 \pi^2 \tilde{l}^{2d-4}}\text{Re }\left(\Theta\left(s,u,v,0\right)-\Theta\left(s,u,v,\tilde{\theta}\right)\right).
\end{equation}
This way the previous example can be rewritten as:
\be
\label{eq.i2b}
I(t_0) = \int_0^1 dx \int_0^\infty dz\,    F_c \left(2,z+2x,x,\tilde{\theta}\right).
\ee
The same can be done will all the terms that appear at order $\lambda_0^2$. For example:
\be
\ec_{1}= \frac{3(1-d)\lambda_0^2 }{2 \vef} \int_0^1 dx \int_0^\infty dz\,  z \, F_c \left(2+xz ,2+(1-x)z,1,\tilde{\theta}\right).  
\ee

\section{Regularisation}

Several of the expressions in the expansion of $\langle E \rangle$ are divergent, and thus require regularisation. 
We will use dimensional regularisation, in a way to be specified below, by setting $d=4-2\epsilon$.

The first step is to identify where these divergences occur. Since the matrix $A(s,u,v,\tilde{\theta})$ is symmetric and has a positive definite real part if 
$\det A(s,u,v,0)\ne 0$, the sum over momenta in $F_c$
will be convergent unless the determinant  at $\thetat =0$ vanishes. For our integrals, that happens for either $u=v=0$ or $s=u=v=2$. Only the first case matters, as a shift in momentum within the sums can bring the latter case to the former. In order to make this more visible, we used Poisson summation to rewrite the theta function:
\begin{equation}
\Theta\left(0\vert iA\left(s,u,v,\tilde{\theta}\right)\right)=(\hc u)^{-\nicefrac{d}{2}}\sum_m \exp\left\{ -\pi\hc s m^{2}\right\} 
\sum_n \left\{ -\frac{\pi}{\hc u} (n-i\hc v m - \thetat \tilde{\epsilon} m)^{2}\right\} .
\end{equation}
And then it is immediate that the $u=v=0$ divergences occur for:
\begin{itemize}
\item The terms in $n=0$ for $\tilde{\theta}=0$.
\item The terms in $n=0$ for which $m_\mu= 0 (\text{ mod } l_{g}), \, \forall \mu$ in a twisted plane, even if $\tilde{\theta}\neq0$.
\end{itemize}
We may then isolate these two divergent contributions. We define:
\be
H\left(s,u,v,\tilde{\theta}\right)= \frac{\hc^2}{16 \pi^2 \tilde{l}^{2d-4}} \sum_n \sum^\prime_m \text{Re } \exp\left\{ -\pi\hc (s m^{2}
+ u n^{2}+2 v m\cdot n) + 2i \thetat m\cdot \tilde{\epsilon}\cdot n)\right\},
\ee
which is automatically finite for $\thetat \ne 0$ (notice the prime in the sum over $m$). We then rewrite $F_c$ as:
\be
F_{c}\left(s,u,v,\thetat \right)=  H\left(s,u,v,0\right)-H\left(s,u,v,\thetat \right) .
\ee
All the divergences are contained in the term $H\left(s,u,v,0\right)$, which can be rendered finite by subtracting:
\be
\label{eq.i3}
H^{\text{div}}\left(s,u,v\right)= \frac{\hat c^2}{16 \pi^2 \tilde{l}^{2d-4}} 
\left(\hat{c}u\right)^{-\frac{d}{2}}\sum^\prime_m \exp\left\{ -\pi\hat{c}\frac{su-v^{2}}{u}m^{2}\right\} .
\ee

One of the most important tests to our calculation is to reproduce the universal  
coefficient of the $1/\epsilon$ term in the relation between bare and renormalised couplings.
The divergent piece of $F_c$ can be computed in dimensional regularisation by rewriting it in terms of Jacobi theta functions:
\be
F_{c}^{\text{div}}\left(s,u,v\right)=\frac{\hc^{2-\nicefrac{d}{2}}}{16\pi^{2}\tilde{l}^{2d-4} u ^{\nicefrac{d}{2}}}
\theta_{3}^{d-d_t}\left(0, \alpha\right)
\left\{ \theta_{3}^{d_t}\left(0, \alpha\right)-\theta_{3}^{d_t}\left(0, l_g^2\alpha\right)\right\} ;\quad \alpha=i\hc\frac{su-v^2}{u}.
\ee
Let us consider $I$ in Eq. (\ref{eq.i2b}). Changing variables 
 $z \rightarrow z/x$, and using the duality relations of $\theta_{3}$:
\be
I (t_0) =  \frac{\hc^{(2-d)}} {16 \pi^2 \tilde l^{(2d-4)}} \Big (  (2\hc)^{\nicefrac{d}{2}}\sum_m^\prime e^{- 2 \pi m^2 \hc }\Big )
\int_0^1 dx \int_0^\infty dz \, x^{1-\frac{d}{2}} (4 + 2z-x)^{-\frac{d}{2}}
+ {\rm finite \, terms} .
\ee
This can be compared with the corresponding expression appearing in infinite volume:
\be
I^\infty (t_0)= \frac {N^2 -1}{N^2}  t_0^{(2-d)} (4 \pi)^{-d} \int_0^1 dx \int_0^\infty dz \, x^{1-\frac{d}{2}} (4 + 2z-x)^{-\frac{d}{2}}.
\ee
Setting $t_0 = \hc \lef^2 / (4\pi)$, one easily derives that:
\be
I (t_0) = {\cal A}_c    \, I^\infty (t_0) + {\rm finite \, terms}, \qquad  {\cal A}_c = \frac {N^2} {N^2 -1} \Big (  (2\hc)^{\nicefrac{d}{2}}  \sum_m^\prime e^{- 2 \pi m^2 \hc }\Big ).
\ee  
The same relation holds for all terms at order $\lambda_0^2$. Similarly, for the leading term in $\lambda_0$: 
\be
\ec_0^{(0)} (t_0) = \frac {\lambda_0 (d-1)}{2 \vef} \sum_m^\prime e^{- 2 \pi m^2 \hc }= {\cal A}_c    \,  \ec_0^{(0)}(t_0)\Big |_{\infty \ {\rm vol}}.
\ee
This implies, in particular, that
\be
\left \langle \frac{E(t_0)}{N} \right \rangle\Big |_{\rm TBC}  = {\cal A}_c    \, 
\left \langle \frac{E(t_0)}{N} \right \rangle\Big |_{\infty \ {\rm vol}} + {\rm finite \, terms},
\ee
and, using the results in \cite{Luscher:2010iy}, 
\be
\left \langle \frac{E(t_0)}{N} \right \rangle\Big |_{\rm TBC}  =  \frac {\lambda_0 (d-1)}{2 \vef}  
\Big (\sum_m^\prime e^{- 2 \pi m^2 \hc }\Big ) \, \Big \{ 1 + \left (\frac{11}{48 \pi^2 \epsilon } + \alpha \right )\lambda_0 + 
{\cal O} (\lambda_0^2 )\Big \}  ,
\ee
with $\alpha$ finite, as we wanted to prove.

Finally, what remains in order to compute the $\Lambda$ parameter in this scheme is to determine the value of the finite pieces. 
There are two contributions to compute: the terms included in $F_{c}$ after subtracting $H^{\text{div}}$, given by:
\be
F_{c}^{\text{fin}}\left(s,u,v,\tilde{\theta}\right)= H\left(s,u,v,0\right)-H^{\text{div}}\left(s,u,v\right)-H\left(s,u,v,\tilde{\theta}\right),
\ee
and the finite parts of $F_{c}^{\text{div}}$.
The latter can be easily treated with algebraic programs such as Mathematica.
As for the former, it is a combination of several multiple integrals in flow time of Siegel theta functions, for whose computation we have prepared a numerical code in C++. 

The code generates the Siegel theta function at each point by summing over increasing values of the momenta in the sum until convergence is reached (defined as a variation of less than 0.1\% with respect to the last value when adding a new order in momenta). It uses the trapezoidal rule to evaluate the flow time integrals up to the desired precision (0.01\% in the current tests). 
For integrals whose upper limit goes to infinity, the integration is separated into intervals of length $\Delta t=1$. Trapezoid integration is then performed for each interval starting from the lower bound until the last interval's contribution represents less than 0.01\% of the total.
Nevertheless, as the computations are still ongoing, we will present the combined results elsewhere.

\section{Summary}

In order to obtain the running of the 't Hooft coupling constant in the twisted gradient flow scheme, we studied the action density up to fourth order in perturbation theory. 
To that goal, we expanded the flow fields in powers of the coupling, solved the flow equations, and rewrote the expectation value of the result in terms of Siegel theta functions. 

This formulation allowed us to identify the terms from which UV divergences were coming, terms which we subsequently subtracted from the observable so as to isolate the finite part. This would allow us to determine the $\Lambda$ parameters between this scheme and the $\overline{\rm MS}$ one in dimensional regularisation. However, the numerical computations are still underway, and will be published later on.

While the computation of the finite part is our main concern, the diverging terms are also very relevant, as their study allowed us to reproduce the universal $1/\epsilon$ term in the relation between bare and renormalised couplings as a consistency check to the validity of our scheme.

Last but not least, the hypothesis of volume independence is also being tested.
Assuming $\tilde{\theta}=\nicefrac{\bar{k}}{l_g}$ is fixed, everything in the previous formulae is given in terms of $\lef$. Note, however, that the presence of the prime in the momentum sums introduces an explicit $N$  dependence in our expressions, an observation already made 
in the context of the TEK one-site model \cite{margaprocs}. As an example, consider the sum appearing 
in ${\cal A}_c$. In $d=4$ it can be expressed as:
\be
\sum_m^\prime e^{- 2 \pi m^2 \hc } =  \sum_{m \ne 0} e^{- 2 \pi \hc m^2 }  -   \sum_{m \ne 0} e^{- 2 \pi \hc N m^2}  .
\ee
The explicit $N$ dependence is thus exponentially suppressed with $N$, which should lead to small corrections to the volume independence conjecture. 
Gauge and spacetime degrees of freedom would thus be linked, and would be redundant for large values of N. 
Nevertheless, this requires further testing; and other factors need to be taken into account, such as the deviation stemming from being unable to keep $\tilde{\theta}$ constant while changing $N$; which will be studied in the future.

\section*{Acknowledgements}

We gratefully acknowledge discussions related to this work with Antonio González-Arroyo and Alberto Ramos.
We acknowledge financial support from the MCINN grants FPA2012-31686
and FPA2015-68541-P, and the Spanish MINECO\textquoteright s \textquotedblleft Centro
de Excelencia Severo Ochoa\textquotedblright{} Programme under grant
SEV- 2012-0249. EIB additionally acknowledges support from the Spanish MINECO's FPI Severo Ochoa grant BES-2015-071791, cofinanced by the European Union's ESF.
The numerical simulations were done on the HPC-clusters
at IFT.

\end{document}